\documentclass[11pt,preprint]{aastex}

\shorttitle{FUV Images of the Vela SNR}
\shortauthors{Nishikida et al.}

\begin{document}

\title{Far Ultraviolet Spectral Images of the Vela Supernova Remnant}

\author{K. Nishikida\altaffilmark{1}, J. Edelstein\altaffilmark{1}, E. J. Korpela\altaffilmark{1}, 
R. Sankrit\altaffilmark{1}, W. M. Feuerstein\altaffilmark{1}, K. W. Min\altaffilmark{2}, 
J-H. Shinn\altaffilmark{2}, D-H. Lee\altaffilmark{2},I-S. Yuk\altaffilmark{3}, H. Jin\altaffilmark{3}, 
K-I. Seon\altaffilmark{3}}

\altaffiltext{1}{Space Sciences Laboratory, University of California, 
Berkeley, CA 94720} 
\altaffiltext{2}{Korea Advanced Institute of Science and Technology, 
305-701, Daejeon, Korea}
\altaffiltext{3}{Korea Astronomy and Space Science Institute, 305-348, 
Daejeon, Korea}

\begin{abstract}
We present far-ultraviolet (FUV) spectral-imaging observations of the Vela supernova remnant (SNR), obtained with the Spectroscopy of Plasma Evolution from Astrophysical Radiation (\emph{SPEAR}) instrument, also known as \emph{FIMS}.  The Vela SNR extends $\sim8^{\circ}$ in the FUV and its global spectra are dominated by shock-induced emission lines.  We find that the global FUV line luminosities can exceed the 0.1--2.5 keV soft X-ray luminosity by an order of magnitude.   The global O~{\small VI}:C~{\small III} ratio shows that the Vela SNR has a relatively large fraction of slower shocks compared with the Cygnus Loop.
\end{abstract}


\section{Introduction}

The Vela supernova remnant (SNR) has been studied in great detail due to its proximity \citep[250pc;][]{vela_distance} and its large angular diameter of $8^{\circ}$ \citep[]{VelaROSAT}.  The remnant is $\sim$10,000 years old \citep{vela_age} and its overall emission is dominated by the interaction of the SN blast wave with the interstellar medium (ISM), but also has a pulsar and plerionic nebula near its center.  

Vela is one of two Galactic SNRs (the other being the Cygnus Loop) that has been extensively studied in the UV.  The UV emission from SNRs arises primarily in shocks driven by the supernova blast wave into interstellar clouds.  The shock velocities responsible for the UV emission lie in the range $\sim$50-300 km s$^{-1}$, and heat the gas to temperatures of ~$10^5 - 10^6$ K.  The hot shocked gas can also be studied via absorption so long as there are suitable background continuum sources.

Absorption line studies of Vela using \textit{Copernicus} showed the presence of O~{\small VI} and N~{\small V}, high ionization species expected in shocked gas, and also high velocity components of lower ionization species \citep{Jenkins1976a,Jenkins1976b}.  Absorption line studies have also been carried out with \textit{IUE} \citep{Jenkins,Nichols}, \emph{HST} \citep{Jenkins1995,Jenkins1998} and \textit{FUSE} \citep{Slavin}.  These studies have shown the existence of fast shocks ($\gtrsim$160 km s$^{-1}$) distributed widely but inhomogeneously across the face of the remnant, and also the variations in the dynamic pressure driving these shocks.

Emission line studies of Vela have been carried out using \textit{IUE} \citep{Vela_IUE}, \textit{HUT} \citep{RaymondVela}, \textit{FUSE} \citep{ravi2,ravi} and \emph{Voyager 2} UVS \citep[henceforth BVL]{vela_voyager}.  These observations, except for the ones obtained by \emph{Voyager 2}, were of regions with angular extents of order an arc-minute and probed the properties
of individual shock fronts.  BVL analyzed \emph{Voyager 2} spectra of a $1.5^{\circ} \times 2.0^{\circ}$ region in the northern part of Vela, which showed strong C~{\small III} and O~{\small VI} emission features.  They found variations in the flux ratios on scales of 0.2$^{\circ}$.  They also concluded from the overall flux ratio between the two lines that slower shocks (those unable to produce O~{\small VI}) were more prevalent in Vela than in the Cygnus Loop.  \textit{FUSE} spectra of a few regions in Vela well separated from each other have strong C~{\small III} lines and relatively weak or no O~{\small VI}, and show the presence of slower shocks ($100 - 140$ km s$^{-1}$)  spread over the face of the remnant \citep{raviproc}.

We present spectral images of the Vela SNR in several FUV lines and FUV spectra of the entire remnant.  The data were obtained with \emph{SPEAR} (The Spectroscopy of Plasma Emission from Astrophysical Radiation), also known as \emph{FIMS} (Far-ultraviolet Imaging Spectrograph).  \emph{SPEAR}, launched Sepember 27, 2003 on the Korean satellite STSAT-1, is a dual-channel FUV imaging spectrograph (S channel 900 - 1150 \AA, L channel 1350 - 1750 \AA, $\lambda/\Delta\lambda\sim$550) with a large imaged field of view (S: 4.0$^{\circ} \times 4.6'$, L: 7.5$^{\circ} \times 4.3'$, spatial resolution $\sim10'$) optimized for the observation of diffuse emission \citep[see][for an overview of the instrument and mission.]{Instrument,Mission}.  A large effective field of view can be obtained by sweeping across the sky.  This combination of instrument properties yields a dataset that supplements data obtained in previous UV studies of the remnant.  The data allow us to estimate the total flux from Vela in several FUV lines.

In the following sections we present the observations (\S2), discuss the spectral images (\S3) and the total spectrum (\S4).  The last section (\S5) summarizes the importance of these data for the study of
SNRs.

\section{Observation and Data Analysis}

The Vela SNR region was observed between January 31 and February 4, 2004.  The data were processed as described in \citet[]{Instrument,Mission} including rejection of data for which the attitude knowledge was poor ($>$30') or data contaminated by airglow, evident from an increased count rate at the end of each orbit ($\leq$10\% of the data).  The resulting number of photons and exposure obtained over 16 orbits were (4.4$\times10^{6}$, 7214 s) and  (1.8$\times10^{5}$, 6307 s) for the L channel and S channel, respectively.  While the SNR region was entirely observed in the L channel,  the S channel coverage was incomplete. 
The photon's sky coordinates ($\alpha$, $\delta$) and wavelengths ($\lambda$) were binned to 0.15$^{\circ}$ (similar to the \emph{SPEAR} imaging resolution after attitude reconstruction) and to 1 \AA, respectively and combined with exposure maps to create a 3-d count-rate data cube ($\alpha$, $\delta$, $\lambda$) for each spectral channel.   We identified data contaminated by bright stars in each channel as pixels in the wavelength-integrated total count rate image that exceeded 3 times the median count rate of a relatively star-free area.  The L and S channels contained 20\% and 10\% of the total pixels identified as stars, respectively.

To create spectral images, the data cube was summed across the waveband of interest with star pixels removed. Star pixel ``holes'' were filled with a linear fit to adjacent pixels at the ``hole's'' declination.  The undetected stars have a flux $<1.5\times 10^{-11}$ ergs s$^{-1}$ cm$^{-2}$\AA$^{-1}$ in the \emph{SPEAR} L channel.  Improved stellar identification, removal and reconstruction methods are currently being developed. 

Net spectral images for specific emission lines, $I_{\lambda_{Net}}$, were constructed by subtracting a continuum image, $I_{cont}$, made from a similar width spectral region \emph{adjacent} to the emission line from an image, $I_{\lambda}$, made from a spectral region \emph{including} the emission line, i.e. $I_{\lambda_{Net}}=I_{\lambda}- I_{cont}$.  This approach often results in an over-subtraction of the lower intensity pixels.  The continuum selected from the global spectra is not suitable for lower intensity pixels because of the non-zero slope of the adjacent continuum.  Therefore we apply a variable scaling factor, $f$, to the subtraction of the continuum image: $I_{\lambda_{Net}} = I_{\lambda}-f \times I_{cont}$.  For each line image, the factor $f$ was set for each image pixel associated with a bin of a five-bin intensity histogram of $I_{\lambda}$ such that 70\% (i.e. $\sim$2$\sigma$) of those pixels would have a net positive flux.  The resulting values of $f$ were 1.0 at high to medium intensity bins and decreased for the lowest one to three intensity bins, depending on $\lambda$, with a typically minimum value of $f=$0.6 for the lowest intensity bin.  The resulting $I_{\lambda_{Net}}$ were smoothed using a 3-pixel (0.45$^{\circ}$) square median smoothing function.  Diffuse spectra were derived by totaling photons and exposure over regions of interest, excluding bright star pixels, and then smoothed by a 3 \AA\ boxcar function.  

\section{Spectral Images}

\begin{figure}
\plotone{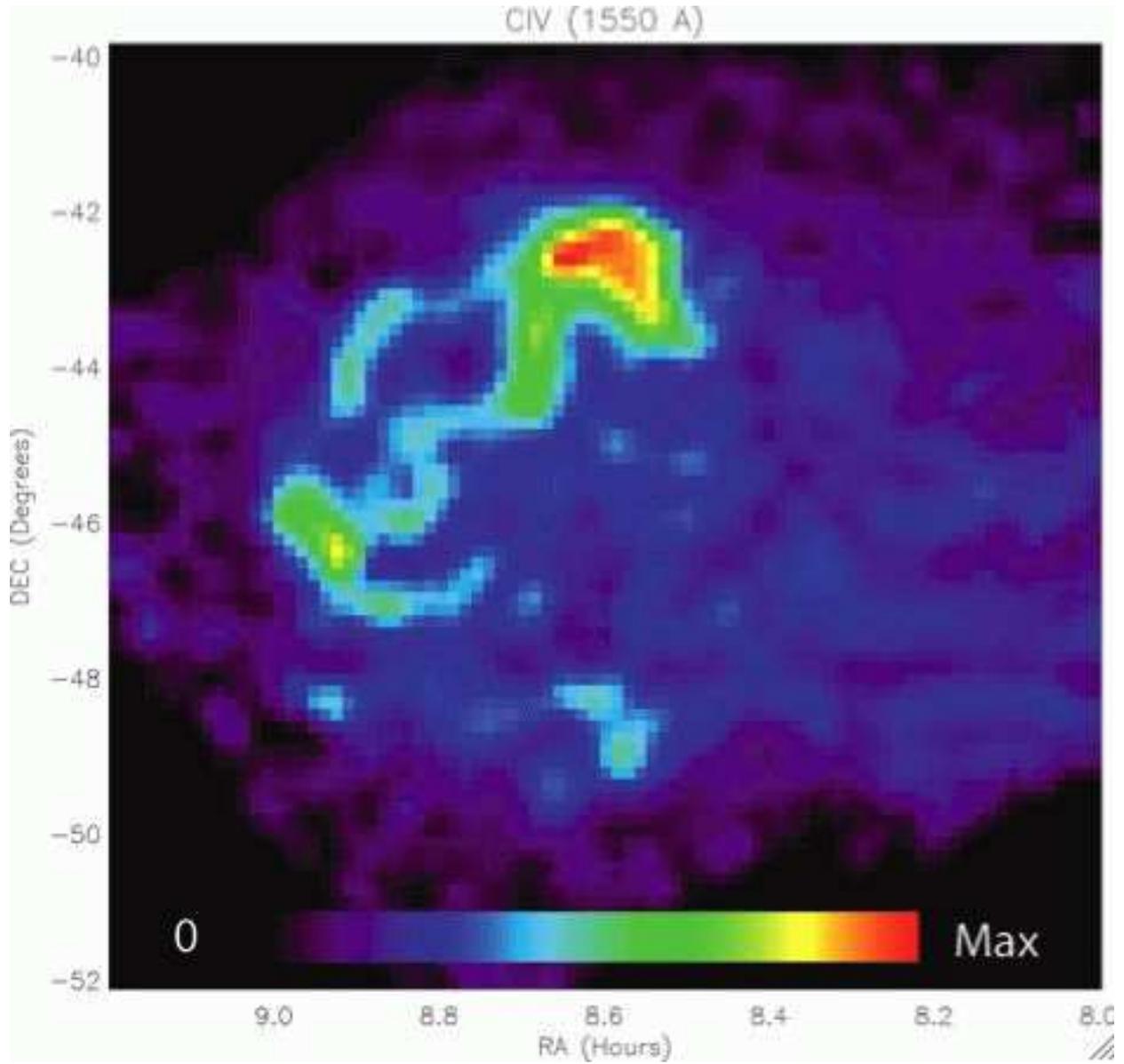}
\caption{\emph{SPEAR} C {\small IV} image of the Vela SNR.  Intensity scale (linear), which applies to the \emph{SPEAR} images in Fig.~\ref{lineimages2} as well, is also shown.  Maximum intensity of the C {\small IV} image is $1.7\times 10^{6}$ LU.}\label{c4image}
\end{figure}

\begin{figure}
\plotone{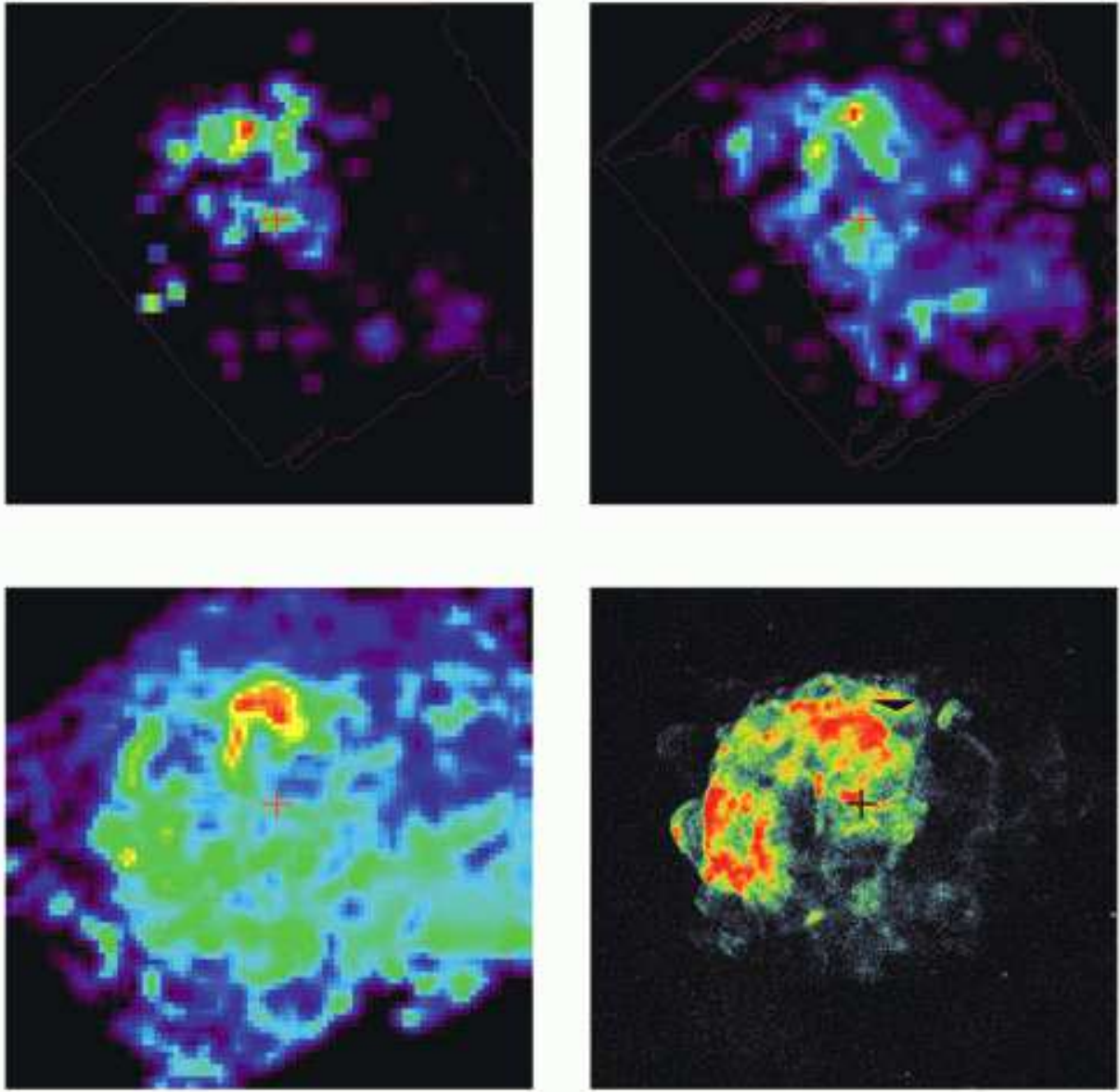}
\caption{Top row: \emph{SPEAR} O {\small VI} (left) and C {\small III} (right) images.  Observation boundary is shown in both images.  The inner contour shown in the C {\small III} image corresponds to 15\% of the maximum exposure in the S channel.  Bottom row: Si {\small IV} \& O {\small IV} (left) and RASS 1/4 keV (right) images.  The ``+'' symbol indicates the position of the Vela pulsar.  Maximum intensities (in $10^{6}$ LU) seen in the O {\small VI}, C {\small III}, and Si {\small IV} \&O {\small IV]} images are 2.9, 4.1, and 0.62, respectively. }\label{lineimages2}
\end{figure}

We show the C {\small IV} (1543--1557 \AA) image of Vela in Fig.~\ref{c4image}.  This is the most prominent emission line in the L channel, and the figure shows the overall morphology of the FUV emission from the SNR.  Images of Vela in C {\small III} (974-980 \AA), O {\small VI} (1019-1033 \AA), Si {\small IV} \& O {\small IV]} (1398--1413 \AA) are shown in Fig.~\ref{lineimages2}.   Also shown for comparison is the \emph{ROSAT} All-Sky Survey (RASS) 1/4 keV image of the same region.  The \emph{SPEAR} maps show that FUV emission is present over most of the Vela SNR.  The C~{\small III} map in particular shows the widespread presence of radiative shocks.  O {\small VI} and C {\small III} trace gas at very different ionization states.  O~{\small VI} production requires a shock velocity of about 150 km/s while C~{\small III}  is produced even in 80 km/s shocks.  There is some overall correspondence between the two, showing where the FUV producing shocks exist, but C~{\small III} is more extended. C~{\small IV} and Si~{\small IV} \& O~{\small IV]} images show a close correlation: C~{\small IV}, Si~{\small IV}, and O~{\small IV]} have comparable ionization potentials and are roughly co-extensive in the post-shock gas.  

Although the FUV and X-ray emission features are not closely correlated, their extent is roughly the same.  \citet{lu} attribute the X-ray emission to thermal emission from a hot, thin gas in the SNR interior.  We used the plasma temperatures and emission measures in Table 1 of \citet{lu} as input for CHIANTI \citep{chianti0,chianti} and confirmed that they produce insufficient thermal FUV emission by several orders of magnitudes compared with the observed emission.  The FUV emission comes from shocks that have been driven into interstellar clouds.  These clouds
are large enough that they have not been destroyed by the blast wave sweeping over them, and they have a high covering factor.  The total FUV luminosity of Vela exceeds the total soft X-ray luminosity (see
\S4).  Thus, the overall radiation rate of the SNR is dominated by shocks in higher density regions traced by their FUV emission.  

The most prominent localized feature is an intense knot of emission near ($\alpha$, $\delta$)=(8.6 h, -42.5$^{\circ}$) that appears at all FUV wavelengths, including the soft X-ray band.  A detailed comparison shows that the peak emission for each FUV wavelength and the soft X-ray are non-coincident and arranged in an arc, suggesting that the feature could be a complex and intense bow shock perhaps analogous to the X-ray ``bullet shocks,'' or ``knots,'' of similar scale.  \citet{VelaROSAT} identified six (labeled A through F) of these X-ray knots and suggested that they were created by high-density ejecta shock-heating the ambient medium.  
The FUV images appear to be limb brightened along the north-east shell boundary and, notably, joins the X-ray knot D to the main shell in C {\small IV} and Si {\small IV}/O {\small IV]}.

FUV emission is absent from the knot B 
region despite its similarity in X-ray intensity to knot D.  This is consistent with FUV emission arising from shocked interstellar media because the knot D bullet is believed to be encountering a more dense ambient media than knot B.  An FUV enhancement particularly notable in O {\small VI}, whose localized intensity peak is coincident with the Vela pulsar, and C {\small III}, which coincides with the Vela-X radio continuum nebula \citep[Figures 1 \& 2]{vela_480,vela_843}, is coincident with a soft X-ray ridge, suggestive of a shock structure.  Curiously, there is little associated C {\small IV} or Si {\small IV} \& O {\small IV]} emission from the same area in comparison to the region of maximum intensity ($\alpha$, $\delta$)=(8.6 h, -42.5$^{\circ}$). This suggests that we are  detecting at  least two markedly different physical conditions in the region surrounding the Vela pulsar.  A more detailed examination will be necessary to tell if these are related, or merely happen to lie along the same line of sight.

\section{FUV Spectra}

\begin{figure}
\plottwo{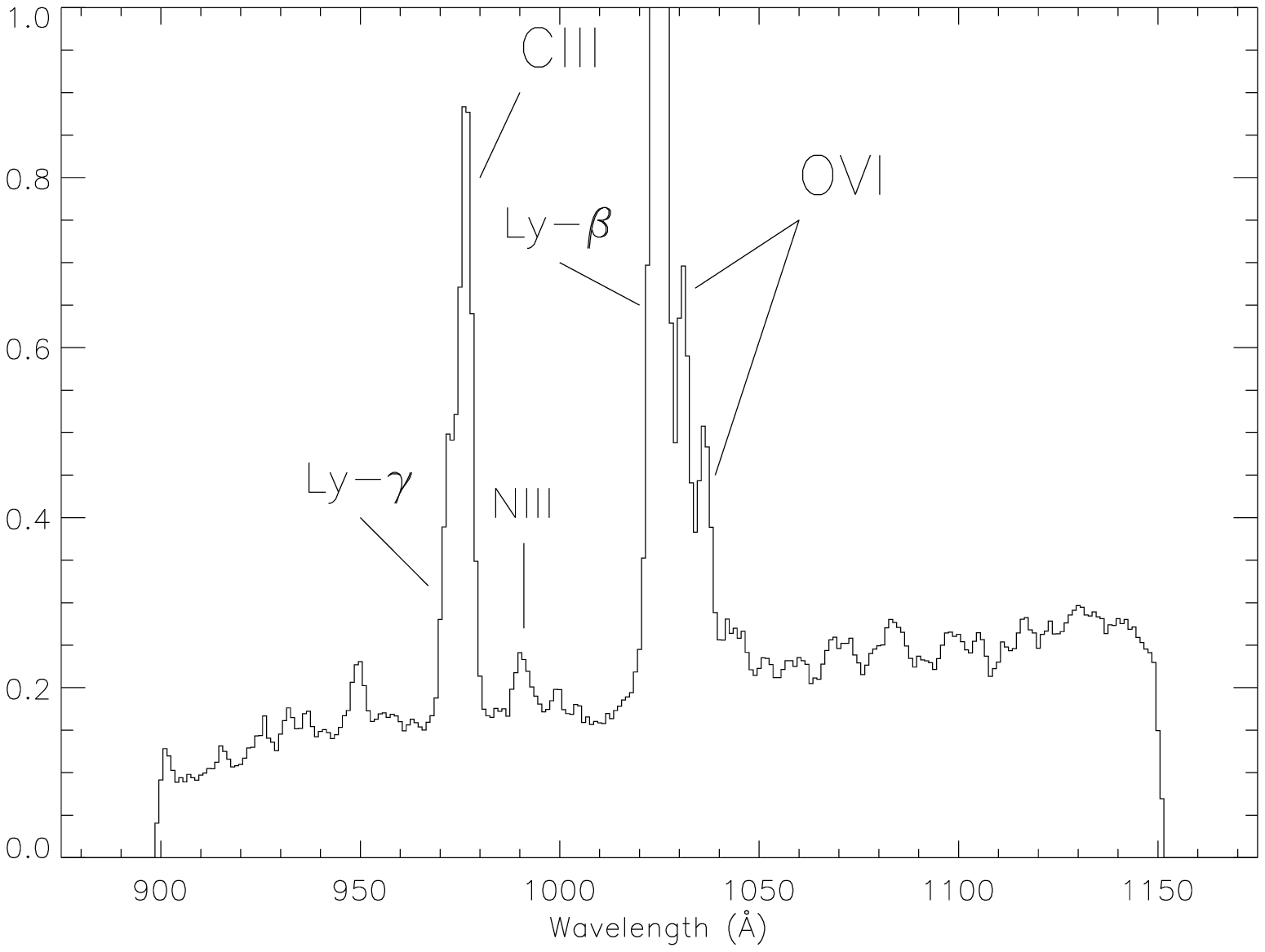}{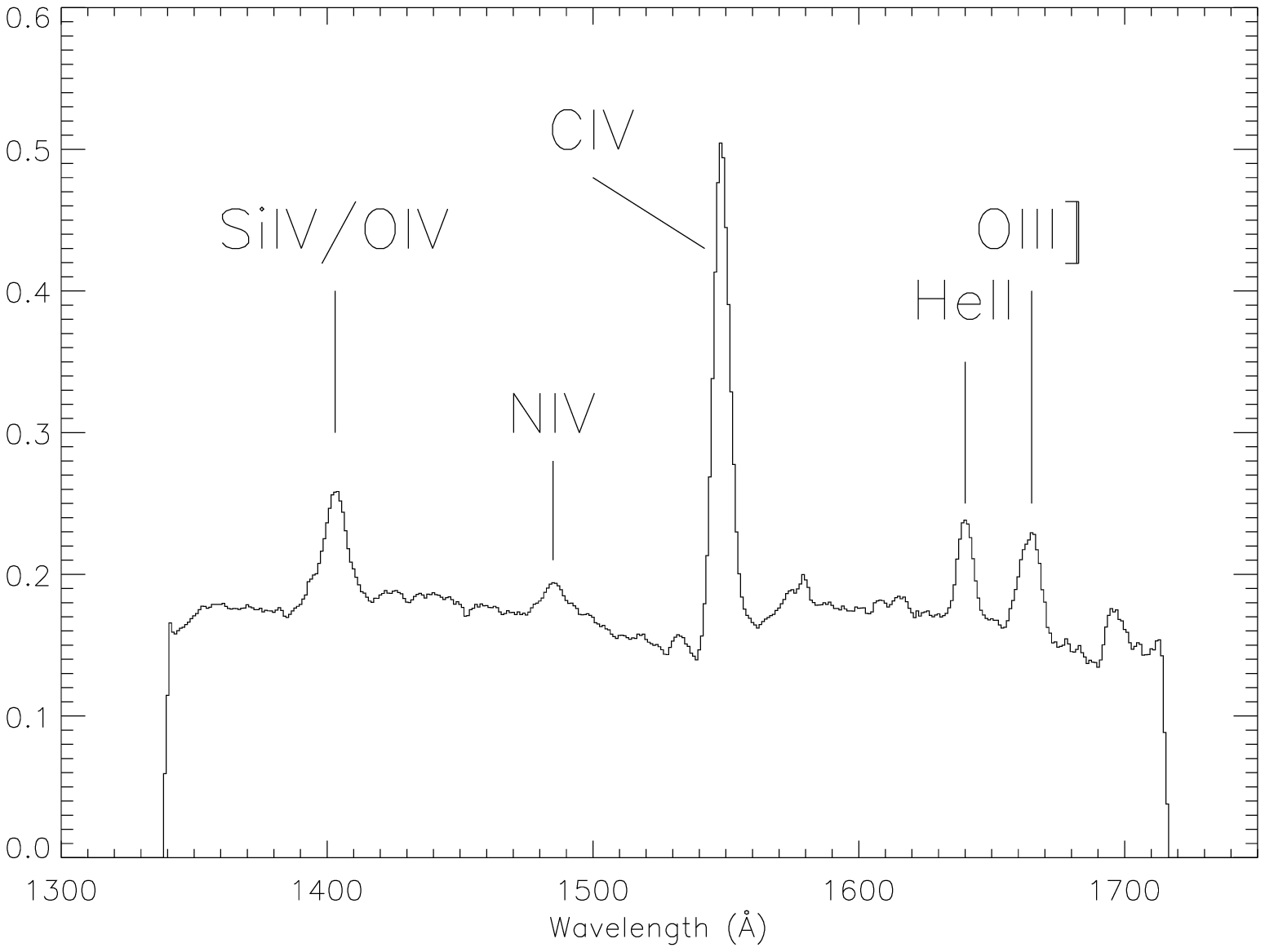}
\caption{The average Vela spectra as obtained with \emph{SPEAR}.  Vertical axis is in units of $10^{5}$ photons s$^{-1}$ cm$^{-2}$ sr$^{-1}$ \AA$^{-1}$.  In addition to Ly-$\beta$ and Ly-$\gamma$, the higher  hydrogen Lyman series can be seen between 910 \AA\ and 950 \AA.  The Lyman series is due to geocoronal emission.}\label{spectra}
\end{figure}

The diffuse \emph{SPEAR} FUV spectra from the entire Vela SNR region are shown in Fig.~\ref{spectra}.  The spectra show strong emission lines from highly ionized atoms produced by high velocity shocks and/or hot plasmas.   Detected lines include C {\small III} (977 \AA), N~{\small III} (990 \AA), O~{\small VI} (1032, 1038 \AA), O {\small IV]} and Si {\small IV} (1400, 1403 \AA\ unresolved), N~{\small IV} (1486 \AA), C {\small IV} (1548, 1550 \AA), He {\small II} (1640 \AA), and O {\small III]} (1660, 1666 \AA).  The emission line-like feature near 1695 \AA\ is instrumental. 

Table~\ref{luminosity} shows the observed FUV line intensities averaged over the SNR.  The C {\small IV}, He {\small II}, and O {\small III]} line profiles were fitted with a Gaussian line profile and a local linear background, while the C {\small III} and O~{\small VI} 1032 \AA\ lines required an additional Gaussian line profile to fit airglow lines adjacent  to C~{\small III} and O {\small VI} $\lambda$1032.  The O {\small VI} doublet  intensity was calculated by multiplying the 1032 \AA\ line intensity by 1.5 (assuming a 2:1 ratio between $\lambda \lambda$1032, 1038) since \emph{SPEAR} does not have the spectral resolution necessary to resolve O {\small VI} $\lambda$1038 and C {\small II}$^{*}$ $\lambda$1037.  For C~{\small III} and O~{\small VI}, we assumed that the average intensity calculated from areas with sky coverage applied to the entire SNR.  Since the Vela SNR appears to be faint at the edges, it is likely that the intensities are slightly overestimated.
The estimated systematic uncertainty in the \emph{SPEAR} sensitivity is $\sim25\%$ \citep{Instrument}.  The \emph{Voyager 2} UVS measurements of O {\small VI} and C {\small III} line intensities at 28 \AA\ resolution (BVL) are in agreement within a factor of two below \emph{SPEAR} measurements in an area near ($\pm1^{\circ}$) UVS pointings.  The global observed O~{\small VI}:C~{\small III} ratio is $\sim1$, consistent within a factor of about two (both above and below) reported by BVL.  

We apply multiplicative correction factors (shown in Table~\ref{luminosity}) to the observed global FUV luminosities, presuming a diameter of $8^{\circ}$ and distance of 250 pc,  to derive dereddened luminosities.  These correction factors were calculated by \citet{ravi} by assuming R=3.1 for selective extinction, E(B-V)=0.1 \citep{Vela_color}, and using the extinction curve suggested by \citet{Vela_ExtinctionCurve}.  The integrated C~{\small III} or O~{\small VI} line luminosity can exceed the entire 0.1 - 2.5 keV luminosity, $2.2\times 10^{35}$ ergs s$^{-1}$ \citep{lu}, by more than an order of magnitude, confirming the importance of the FUV waveband to SNR cooling. 

Vela's global FUV luminosity can be compared with that of the Cygnus Loop \citep{cygnus_voyager}.
Vela is about five times less luminous than Cygnus in the X-ray and in C {\small IV}, about twice as faint in O {\small VI}, and is equally as luminous in C {\small III}.  The observed  O~{\small VI}:C~{\small III} ratio of Cygnus is $\sim2$, illustrating the relatively large fraction of slower shocks in Vela compared with Cygnus. For Cygnus, \citet{cygnus_rocket} showed that the O {\small VI} luminosity is at least as much as the 0.1--4 keV X-ray luminosity \citep{cygnus_einstein}; the combined O~{\small VI}, C~{\small III}, C~{\small IV} luminosity was found to be ten times as large as the X-ray luminosity \citep{cygnus_voyager}.  
In both the Cygnus Loop and Vela, the supernova blast wave is expanding into inhomogeneous surroundings, and the radiative shocks in the denser interaction regions emit strongly in the FUV.

\section{Summary and Future Work}
We have presented global FUV spectral images and spectra of the Vela SNR.  Our data indicate inhomogeneous shock-induced emission from the SNR surface.  The images show limb brightening and knots of emission.  The spectra are consistent with past FUV observations of Vela. The global FUV luminosities of emission lines can exceed the soft X-ray luminosity by an order of magnitude, providing an efficient cooling channel to the SNR.  

The data will be used to compare the global FUV morphology with images in other wavebands and examine how the SNR evolves and interacts with the ambient ISM.  \emph{SPEAR} spectra toward previously observed regions \citep[ex.][]{RaymondVela,ravi2} will be modeled in detail to determine the physical properties of the regions.  Mapping the O~{\small VI}:C~{\small III} ratio will allow us to map the distribution of shock velocities across Vela.  Furthermore, the global \emph{SPEAR} spectra can be compared with those of neighboring regions to investigate Vela's association with the Gum nebula.

\acknowledgements
\emph{SPEAR / FIMS} is a joint project of KASSI \& KAIST (Korea) and U.C., Berkeley (USA), funded by the Korea MOST and NASA Grant NAG5-5355.  We used NASA's \emph{SkyView} (http://skyview.gsfc.nasa.gov) facility and CHIANTI, a collaboration of NRL (USA), RAL (UK), U. Florence 
(Italy) and Cambridge (UK).

\clearpage

\begin{table}[htbp]

\caption{Observed global FUV intensities 
and dereddened luminosities of  the Vela SNR}
\begin{tabular}{c|cccccc}
Species & C~{\small III} & O~{\small VI} & C~{\small IV} & He~{\small II} & O~{\small III]} & X-ray \\ \hline\hline
Observed intensity ($10^5 LU$) & 2.6 & 2.7 & 2.3 & 0.5 & 0.5 & \\
Reddening correction & 4.47 & 3.73 & 2.07 & 2.03 & 2.03 &  \\
Dereddened luminosity ($10^{35}$ ergs s$^{-1}$) & 26.8  & 22.7 & 8.0 & 1.3 & 1.5 & 2.2 \\
\end{tabular}

\tablecomments{Diameter of 8$^{\circ}$ and 250 pc distance assumed.  The O~VI luminosity was calculated assuming a 2:1 line ratio between O~VI 1032 \AA\ and 1038 \AA\ lines.  Reddening correction values are taken from \citet{ravi}.  X-ray (0.1--2.5 keV) luminosity from \citet{lu}.  1 Line Unit (LU) = 1 photon s$^{-1}$ cm$^{-2}$ sr$^{-1}$ = 1.9$\times 10^{-11}$ ergs s$^{-1}$ cm$^{-2}$ sr$^{-1}$ at 1032 \AA\ and 1.3$\times 10^{-11}$ ergs s$^{-1}$ cm$^{-2}$ sr$^{-1}$ at 1550 \AA.} \label{luminosity}
\end{table}

\clearpage

\clearpage

\clearpage

\begin{thebibliography}{28}
\expandafter\ifx\csname natexlab\endcsname\relax\def\natexlab#1{#1}\fi

\bibitem[{Aschenbach {et~al.}(1995)Aschenbach, Egger, \& Tr\"umper}]{VelaROSAT}
Aschenbach, B., Egger, R., \& Tr\"umper, J. 1995, Nature, 396, 587

\bibitem[{{Blair} {et~al.}(1991){Blair}, {Long}, {Vancura}, \&
  {Holberg}}]{cygnus_voyager}
{Blair}, W.~P., {Long}, K.~S., {Vancura}, O., \& {Holberg}, J.~B. 1991, \apj,
  374, 202

\bibitem[{Blair {et~al.}(1995)Blair, Vancura, \& Long}]{vela_voyager}
Blair, W.~P., Vancura, O., \& Long, K.~S. 1995, \aj, 110, 312

\bibitem[{{Bock} {et~al.}(1998){Bock}, {Turtle}, \& {Green}}]{vela_843}
{Bock}, D.~C.-J., {Turtle}, A.~J., \& {Green}, A.~J. 1998, \aj, 116, 1886

\bibitem[{Cha {et~al.}(1999)Cha, Sembach, \& Danks}]{vela_distance}
Cha, A.~N., Sembach, K.~R., \& Danks, A.~C. 1999, \apj, 515, L25

\bibitem[{Dere {et~al.}(1997)Dere, Landi, Mason, Monsignori~Fossi, \&
  Young}]{chianti0}
Dere, K.~P., Landi, E., Mason, H.~E., Monsignori~Fossi, B.~C., \& Young, P.~R.
  1997, \apjs, 125, 149

\bibitem[{Edelstein {et~al.}(2005{\natexlab{a}})Edelstein, Korpela, Adlofo,
  Bowen, Feuerstein, Hull, Jelinsky, Nishikida, McKee, Berg, Chung, Fischer,
  Min, Oh, Rhee, Ryu, Shinn, Han, Lee, Seon, Jin, Yuk, Park, \&
  Nam}]{Instrument}
Edelstein, J., Korpela, E.~J., Adlofo, J., Bowen, M., Feuerstein, W.~M., Hull,
  J., Jelinsky, S., Nishikida, K., McKee, K., Berg, P., Chung, R., Fischer, J.,
  Min, K.~W., Oh, S.-H., Rhee, J.-G., Ryu, K., Shinn, J.-H., Han, W., Lee,
  D.-H., Seon, K.-I., Jin, H., Yuk, I.-S., Park, J.-H., \& Nam, U.-W.
  2005{\natexlab{a}}, \apj, submitted

\bibitem[{Edelstein {et~al.}(2005{\natexlab{b}})Edelstein, Min, Han, Korpela,
  Nishikida, Welsh, Heiles, Feuerstein, Adlofo, Bowen, McKee, Lim, Ryu, Shinn,
  Nam, Park, Yuk, Jin, Seon, Lee, \& Sim}]{Mission}
Edelstein, J., Min, K.~W., Han, W., Korpela, E.~J., Nishikida, K., Welsh,
  B.~Y., Heiles, C., Feuerstein, W.~M., Adlofo, J., Bowen, M., McKee, K., Lim,
  J.-T., Ryu, K., Shinn, J.-H., Nam, U.-W., Park, J.-H., Yuk, I.-S., Jin, H.,
  Seon, K.-I., Lee, D.-H., \& Sim, E. 2005{\natexlab{b}}, \apj, submitted

\bibitem[{Fitzpatrick(1999)}]{Vela_ExtinctionCurve}
Fitzpatrick, E.~L. 1999, \pasp, 111, 63

\bibitem[{{Haslam} {et~al.}(1982){Haslam}, {Stoffel}, {Salter}, \&
  {Wilson}}]{vela_480}
{Haslam}, C.~G.~T., {Stoffel}, H., {Salter}, C.~J., \& {Wilson}, W.~E. 1982,
  \aaps, 47, 1

\bibitem[{{Jenkins} {et~al.}(1998){Jenkins}, {Tripp}, {Fitzpatrick}, {Lindler},
  {Danks}, {Beck}, {Bowers}, {Joseph}, {Kaiser}, {Kimble}, {Kraemer},
  {Robinson}, {Timothy}, {Valenti}, \& {Woodgate}}]{Jenkins1998}
{Jenkins}, E.~B., {Tripp}, T.~M., {Fitzpatrick}, E.~L., {Lindler}, D., {Danks},
  A.~C., {Beck}, T.~L., {Bowers}, C.~W., {Joseph}, C.~L., {Kaiser}, M.~E.,
  {Kimble}, R.~A., {Kraemer}, S.~B., {Robinson}, R.~D., {Timothy}, J.~G.,
  {Valenti}, J.~A., \& {Woodgate}, B.~E. 1998, \apjl, 492, L147

\bibitem[{Jenkins \& Wallerstein(1995)}]{Jenkins1995}
Jenkins, E.~B. \& Wallerstein, G. 1995, \apj, 440, 227

\bibitem[{{Jenkins} {et~al.}(1976{\natexlab{a}}){Jenkins}, {Wallerstein}, \&
  {Silk}}]{Jenkins1976a}
{Jenkins}, E.~B., {Wallerstein}, G., \& {Silk}, J. 1976{\natexlab{a}}, \apjl,
  209, L87

\bibitem[{{Jenkins} {et~al.}(1976{\natexlab{b}}){Jenkins}, {Wallerstein}, \&
  {Silk}}]{Jenkins1976b}
---. 1976{\natexlab{b}}, \apjs, 32, 681

\bibitem[{Jenkins {et~al.}(1984)Jenkins, Wallerstein, \& Silk}]{Jenkins}
Jenkins, E.~B., Wallerstein, G., \& Silk, J. 1984, \apj, 278, 649

\bibitem[{{Ku} {et~al.}(1984){Ku}, {Kahn}, {Pisarski}, \&
  {Long}}]{cygnus_einstein}
{Ku}, W.~H.-M., {Kahn}, S.~M., {Pisarski}, R., \& {Long}, K.~S. 1984, \apj,
  278, 615

\bibitem[{Lu \& Aschenbach(2000)}]{lu}
Lu, F.~J. \& Aschenbach, B. 2000, \aap, 362, 1083

\bibitem[{Nichols \& Slavin(2004)}]{Nichols}
Nichols, J.~S. \& Slavin, J.~D. 2004, \apj, 610, 285

\bibitem[{Rasmussen \& Martin(1991)}]{cygnus_rocket}
Rasmussen, A. \& Martin, C. 1991, \apj, 396, 103L

\bibitem[{Raymond {et~al.}(1997)Raymond, Blair, Long, Vancura, Edgar, Morse,
  Hartigan, \& Sanders}]{RaymondVela}
Raymond, J.~C., Blair, W.~P., Long, K.~S., Vancura, O., Edgar, R.~J., Morse,
  J., Hartigan, P., \& Sanders, W.~T. 1997, \apj, 482, 881

\bibitem[{Raymond {et~al.}(1991)Raymond, Wallerstein, \& Balick}]{Vela_IUE}
Raymond, J.~C., Wallerstein, G., \& Balick, B. 1991, \apj, 383, 226

\bibitem[{Reichley {et~al.}(1970)Reichley, Downs, \& Morris}]{vela_age}
Reichley, P.~E., Downs, G.~S., \& Morris, G.~A. 1970, \apj, 159, L35

\bibitem[{Sankrit(2004)}]{raviproc}
Sankrit, R. 2004, in Proceedings of the conference ``How Does The Galaxy
  Work?'', ed. E.~J. Alfaro, E.~Perez, \& J.~Franco (Dordrecht; London; Kluwer
  Academic), 177

\bibitem[{Sankrit {et~al.}(2003)Sankrit, Blair, \& Raymond}]{ravi}
Sankrit, R., Blair, W.~P., \& Raymond, J.~C. 2003, \apj, 587, 242

\bibitem[{Sankrit {et~al.}(2001)Sankrit, Shelton, Blair, Sembach, \&
  Jenkins}]{ravi2}
Sankrit, R., Shelton, R.~L., Blair, W.~P., Sembach, K.~R., \& Jenkins, E.~B.
  2001, \apj, 549, 416

\bibitem[{Slavin {et~al.}(2004)Slavin, Nichols, \& Blair}]{Slavin}
Slavin, J.~D., Nichols, J.~S., \& Blair, W.~P. 2004, \apj, 606, 900

\bibitem[{Wallerstein \& Balick(1990)}]{Vela_color}
Wallerstein, G. \& Balick, B. 1990, \mnras, 245, 701

\bibitem[{Young {et~al.}(2003)Young, Del~Zanna, Landi, Dere, Mason, \&
  Landini}]{chianti}
Young, P.~R., Del~Zanna, G., Landi, E., Dere, K.~P., Mason, H.~E., \& Landini,
  M. 2003, \apjs, 144, 135

\end{thebibliography}
\end{document}